# Genome-scale reconstruction of the metabolic network in *Yersinia pestis*, strain 91001

**Ali Navid and Eivind Almaas**[*]

Biosciences & Biotechnology Division, Lawrence Livermore National Laboratory, Livermore, California 94550-0808, USA

[*]Corresponding author

Email address: almaas@llnl.gov

## Abstract

The gram-negative bacterium *Yersinia pestis*, the aetiological agent of bubonic plague, is one the deadliest pathogens known to man. Despite its historical reputation, plague is a modern disease which annually afflicts thousands of people. Public safety considerations greatly limit clinical experimentation on this organism and thus development of theoretical tools to analyze the capabilities of this pathogen is of utmost importance.

Here, we report the first genome-scale metabolic model of *Yersinia pestis* biovar Mediaevalis based both on its recently annotated genome, and physiological and biochemical data from literature. Our model demonstrates excellent agreement with *Y. pestis*' known metabolic needs and capabilities. Since *Y. pestis* is a meiotrophic organism, we have developed CryptFind, a systematic approach to identify all candidate cryptic genes responsible for known and theoretical meiotrophic phenomena. In addition to uncovering every known cryptic gene for *Y. pestis*, our analysis of the rhamnose fermentation pathway suggests that *betB* is the responsible cryptic gene.

Despite all of our medical advances, we still do not have a vaccine for bubonic plague. Recent discoveries of antibiotic resistant strains of *Yersinia pestis* coupled with the threat of plague being used as a bioterrorism weapon compel us to develop new tools for studying the physiology of this deadly pathogen. Using our theoretical model, we can study the cell's phenotypic behavior under different circumstances and identify metabolic weaknesses which may be harnessed for the development of therapeutics. Additionally, the automatic identification of cryptic genes expands the usage of genomic data for pharmaceutical purposes.

# Introduction

The "black death" pandemic that ravaged Europe between the $14^{th}$ to $16^{th}$ centuries is the most infamous outbreak of pestilence in history. Within a short five-year period (1347-1352), thirty three million people, one out of every three Europeans perished. The Mediaevalis biovar of the gram-negative bacterium *Yersinia pestis* (YP), the aetiological agent of bubonic plague, is believed to have caused this epidemic [1, 2]. The most recent outbreak of bubonic plague in Asia killed nearly 12.5 million people in India alone from 1889 to 1950. Throughout human history, a conservative estimate stipulates that 200 million people have been victims to this deadly disease in various pandemics [3]. Despite modern advances in medicine, no working vaccine against the plague exists, and it is listed by the US Centers for Disease Control and Prevention (CDC) as a Category A bio-terrorism pathogen. While plague is frequently considered a disease of the past, several thousand new cases are reported each year, predominantly in Africa [4]. Between 1990 to 1995, the Democratic Republic of Congo, Tanzania and Zimbabwe alone reported 4939 cases of plague [5]. Hence, the recent reports of antibiotic resistant strains of YP [6-8] are cause for great alarm.

Over the past decade, the revolutionary advances in high-throughput technologies and computational approaches have led to the inception of systems biology, which aims to transform microbiology from a science which focuses on one specific cellular process or pathway, to one that the biology of the system as a whole is examined. To achieve this goal, genome-scale models of metabolism have been developed using constraint-based approaches with flux-balance analysis (FBA) being the most widely used method. FBA's success originates from the fact that, unlike kinetic models, FBA only seeks to identify optimal metabolic steady-state activity patterns that satisfy constraints imposed by mass balance, the metabolic network structure, and the availability of nutrients. The most common objective function (cellular task to be optimized) is that of growth, although other choices are possible depending on the selective environment of the cell [9]. FBA has been applied to several genome-scale models [10-14] with great success. Additionally, FBA has been used to examine a range of topics from the global organization of metabolic fluxes [15] to the effect of genetic knockouts [16] and the discovery of novel regulatory interactions [17]. However, the overall process of system building is still highly labor intensive and requires extensive human curation to generate high-fidelity models.

Here, we present and analyze the first genome-scale reconstruction of an organism classified by the CDC as a Category A pathogen. We demonstrate excellent agreement with known metabolic performance for YP. The model predictions match the identified nutritional requirements and metabolic efficiency of YP, as well as established characteristics for biovar Mediaevalis and strain 91001. Furthermore, we have developed a general method based on FBA for the identification of candidate cryptic and pseudogenes. Applying it to seven meiotrophic pathways of YP, we identify known cryptic genes for the phenylalanine, methionine, and urease pathways in addition to novel cryptic gene candidates for the other pathways.

# Results

### Metabolic network reconstruction

The *Yersinia pestis* model iAN818m is based on the sequenced genome of Mediaevalis strain 91001 [18]. Of the 4037 genes present in the genome, 1146 are believed to be related to cellular metabolism. Our model accounts for the activity of 818 of these genes (71%), resulting in 969 enzymatic reactions. Additional literature surveys identified the activity of 37 local orphan enzymes (13-critical for biomass production, 20-based on literature, 4-pathway hole-filling) and 14 non-enzymatic reactions, resulting in a final model of 1020 reactions and 825 metabolites.

Several studies [19-22] have shown that the composition of YP's cellular membrane changes when the cell transitions from the flea gut environment (high $Ca^{2+}$, 26°C) to that of the mammalian host (low $Ca^{2+}$, 37°C). We have implemented this change in our model by developing two separate biomass compositions. The model includes the pathways for production of yersiniabactin virulence factor; however it currently does not contain the biosynthetic pathways for the production of other pathogenic proteins such as yersinia outer proteins. A summary of the model characteristics is given in Table 1. Figure 1 displays the breakdown of reactions based on metabolic pathway affiliation. This breakdown is similar to that of metabolism in a number of other modeled organisms. A complete list of the metabolic reactions is available in the Supplementary Material.

**Meiotrophy**

In the 1950s, it was realized that *Yersinia pestis* is a meiotrophic organism [23], and thus, its genome is replete with cryptic genes. Cryptic genes are not transcribed by a cell under normal conditions, and are not critical for its proper functioning. However, these genes may be activated through mutations when a cell's environment has been perturbed such that their activity becomes crucial for cellular survival [24, 25]. While cryptic genes do not confer direct (positive) contributions to the fitness of a cell, they serve a crucial role in the survival of a species by endowing the organism with an endogenous genetic reservoir that enhances its adaptive capability [25]. Theoretical studies have shown that the loss or degradation of cryptic genes from a microbial population is very rare, and that these genes may remain in a genome as long as selection for their reactivation occurs occasionally [26].

The activation of cryptic genes during periods of natural selection has been reported in a number of bacteria. For example, the mutational activation of the *bgl* operon enables some organisms to survive on an assortment of β-glucosides [27-29]. In *Escherichia coli,* the mutational reversal of a frameshift in the *ilvG* gene leads to expression of an α-acetohydroxyacid synthase isozyme which is not susceptible to feedback inhibition by valine [30]. Similarly, mutations in the *citA* and *citB* genes of *E. coli* provide the cell with the capability to utilize citrate as the sole carbon source [31]. Since the metabolic potential of cells is enhanced by the activation of cryptic genes, the identification and targeted utilization of these genes has a significant potential for industrial and pharmacological benefits.

In YP, meiotrophy has been observed in the biosynthetic pathways of glycine/threonine [32, 33], L-valine and L-isoleucine [34], L-phenylalanine [32], and L-methionine [23, 32] as well as fermentation of melibiose [35], rhamnose [36], and in the urease pathway [37]. Consequently, our base-reconstruction predicts that none of these substrates are critical for growth since their metabolic capability is ostensibly present in the annotated genome. For melibiose and rhamnose, the base-reconstruction is in agreement with experiments due to the fact that strain 91001 is different from most YP strains by being capable of utilizing these two carbohydrates as carbon sources[18].

To systematically identify candidate cryptic genes responsible for the observed meiotrophy in YP, we have developed the CryptFind approach that screens all available genes (see Methods). The complete list of candidate cryptic genes responsible for observed meiotrophic behavior in YP metabolism is given in Table 2. In agreement with previous studies that have classified *pheA, metB,* and *ureD* as cryptic genes [38], our approach identified these among the few candidate cryptic genes in their respective pathways. Although strain 91001 is melibiose and rhamnose positive [18], we determined a number of candidate genes that in other strains of YP could be cryptic for these pathways. In particular, gene YP1470, which translates into melibiose carrier protein *MelB*, has previously been designated as a cryptic gene [18]. In case of rhamnose, we identify the candidate genes to be *rhaT* (L-rhamnose proton symport protein), *rhaA* (L-rhamnose isomerase; EC 5.3.1.14), *tpiA* (triose phosphate isomerase; EC 5.3.1.1), and *betB* (betaine-aldehyde dehydrogenase; EC. 1.2.1.8). Based on Englesberg's [39] observation that rhamnose positive mutants do not fully oxidize rhamnose and excrete small amounts of lactaldehyde into the medium, we suggest that *betB* is the likely cryptic gene. In response to these findings, we have augmented the model and ensured that the end products of the identified meiotrophic pathways in strain 91001 are inactive when we simulate the normal state (wild type) of our model.

To further emphasize the utility of CryptFind, we have analyzed the arabinose fermentation pathway known to be inactive in strain 91001 [18]. While it is known that meiotrophy is unlikely for this pathway due to a significant frameshift caused by a 122 basepair deletion in *araC* [18], CryptFind may still be used to identify candidate genes responsible for known, or hypothesized, pathway deactivations (see Table 2). We note that pseudogene *araC* is among the short list of candidate genes identified by CryptFind.

**Nutrient requirements and metabolic efficiency**

In order to determine the wild type YP's minimum nutritional requirement, we allowed for the import of all possible sources of carbon, oxygen, nitrogen, phosphate and sulfur. Our results indicate that optimal cellular growth requires a carbon, sulfur and phosphorus source. The sulfur requirement can be satisfied either by sulfate, thiosulfate or L-cysteine. The cell also has an obligate requirement for the amino acids isoleucine, valine, methionine, and either glycine or threonine. The only obligate amino acid that is not part of the metabolic pathways is methionine, which is only imported and directly incorporated into cellular biomass. Consequently, we constrain growth in the YP model by limiting the methionine transport to a maximal rate that corresponds to the experimentally reported doubling time of two hours [40, 41]. Based on these observations, we used a medium composed of gluconate, phosphate, thiosulfate, methionine, phenylalanine, isoleucine, valine and glycine as a “minimal” (MIN) growth medium for the wild type YP simulations. For simulating the cell’s growth in a nutritionally ideal environment, we used the TMH medium [42].

The YP metabolism has previously been characterized as highly inefficient, with only 58% of the carbon imported into the cell as glucose being assimilated into biomass [43]. A suggested reason for this inefficiency is that YP lacks one of the primary means by which bacteria silence their pseudogenes [44]. This deficiency could lead to a significant amount of metabolic capability being wasted on expression and degradation of genes that no longer code for proper functions. To systematically evaluate the metabolic efficiency of iAN818m, we simulated cellular growth for multiple different carbon sources. Since the choice of carbon source and oxygen availability may impact the efficiency calculations, we introduced a uniform maximal carbon source uptake rate corresponding to 30 carbons, e.g. the maximal uptake rate for glucose and glycerol is 5 and 10 mmol/gdw hr, respectively. The maximal uptake rate for oxygen, set to 20 mmol/gdw hr, was never reached. Based on this result, if one assumes that the carbon and oxygen uptake mechanisms of YP have similar capacities to those of other enteropathogens that have much shorter doubling times (such as *E. coli*[11] and YP's progenitor *Yersinia pseudotuberculosis*[40] (YPS)), one can conclude that respiration for YP is not rate limiting. However, if the uptake of oxygen is severely stunted in comparison to other enteropathogens, YP's cellular metabolism would be respiration limited. Due to a lack of experimental measurements as well as YP's close relationship to YPS and *E. coli*, our results are based on the earlier assumption.

Figures 2 and 3 demonstrate that the reconstructed metabolic network also is highly carbon-inefficient, with a large amount of the imported carbon being wasted in all studied conditions. In Figure 3, we show the model-predicted utilization of the central carbon metabolism using either glucose or gluconate as the carbon source in a nutrient rich aerobic medium. In rich media, via deamination, some of the imported amino acids can serve as a carbon source. For example, as shown in Figure 3, deamination of glutamate to α-ketoglutarate is a major source of carbon for the citric acid cycle. Additionally, our simulations of YP metabolism for both nutrient rich and nutrient poor media show that the activity of the enzyme α-ketoglutarate dehydrogenase (EC 1.2.4.2) is either absent or very small when compared to other TCA reactions. This seems to corroborate the prediction that there might be a deficiency in the activity of this enzyme in *Y. pestis* [45].

It is important to note that, as is the case with reported flux solutions for all FBA models, calculated flux values that optimize an objective function typically are not unique. The magnitude of the flux degeneracy has been shown to substantially depend on environmental conditions and network composition[46]. In order to evaluate the impact of degeneracy on our flux distributions, we conducted a flux variability analysis[46] in a fixed environment for the reactions shown in Figure 3. Our simulations show that flux degeneracy is not significant for the majority of the reactions (see Supplementary Material).

We find that the YP metabolism is considerably more carbon efficient in the presence of oxygen than in anaerobic environments (see Figure 2). According to the stoichiometry of YP's oxidative phosphorylation, the P/O ratio for NADH in our model can vary from 0.66 to 2. The stoichiometry of proton import in YP via oxidative phosphorylation is similar to that of *E. coli*: YP, like *E. coli*, possesses both NDH-I and NDH-II types of NADH dehydrogenase enzyme[47]. Furthermore, the cytochrome complement of both organisms consists of cytochrome bd and cytochrome o. Protein BLAST analyses verify that there is a high degree of similarity (ranging from 65% to 86%) between the subunits of cytochromes of YP and E. coli. Based on this resemblance and because of a dearth of studies into the mechanism of oxidative phosphorylation in YP, we have used the stoichiometric coefficients from the electron transport mechanism of a published model of E. coli[11]. The calculated P/O ratios for our simulations are reported in Table 3.

While experiments have shown that the growth of YP is robust immediately following the incubation in a glucose-rich medium; as the metabolic byproducts accumulate, glucose is no longer a suitable carbon source due to rapid acidification of the extracellular medium [34]. Experimentally, this challenge is overcome by the use of gluconate as primary carbon source [48, 49]. As can be seen from Table 3 and Figure 2, the model identifies gluconate as the carbohydrate in nutritionally minimal environments that acidifies the medium the least while maintaining carbon efficiencies comparable to that of glucose. Thus, our model analysis supports the time proven choice of gluconate as the ideal carbon source for batch cultivation of YP. Finally, in accordance with experimental observations in nutrient poor and anaerobic conditions [50], the model predicts that the presence of $CO_2$ and/or bicarbonate potentially improves cellular growth (see Table 3).

**Metabolic robustness**

Genome-scale theoretical models present a powerful tool for the analysis of a cell's strengths and vulnerabilities. While it is becoming a standard procedure to conduct large-scale experimental single-gene knockout screens for organisms such as *E. coli* and *S. cerevisiae*, this remains a highly non-trivial task for Category A pathogens due to restrictions on the use of antibiotics. We have characterized the resistance of YP to possible drug interactions and genetic mutations for all single-gene and double-gene knockouts in multiple environments *in silico*. We systematically identified all essential single reaction (SRKOs) and gene knockouts (SGKOs) as well as synthetic lethal mutations (SLMs) (see Figure 4). We find that in all environments there are 74 critical SGKOs and 77 SRKOs. Under aerobic conditions, the model contains 126 essential SGKOs and 61 SLMs in the TMH medium, while the MIN medium has 168 critical SGKOs and 56 SLMs. For both media, absence of oxygen adds 1 SGKO and 16 SLMs. Additionally, we identify an additional 1 SGKO and 2 SLMs that are solely associated with anaerobic, nutrient poor conditions.

Nutritionally poor media significantly limit the number of ways the metabolic network can incorporate nutrients, and thus, cells growing in these environments are more fragile. The number of growth limiting SGKOs are drastically smaller (>25% reduction) for a rich medium in comparison to a poor medium. A complete list of critical genes is given in the Supplementary Material.

In MIN, over 80% of the extra SGKOs (in comparison to TMH) are involved in the production of amino acids that are obligatory for cellular growth. These amino acids are available in TMH, and the removal of their biosynthetic genes has no effect on cellular growth. Other lethal SGKOs are involved in the production of cell envelope material as well as necessary intermediates of the pentose phosphate pathway. In particular, the evolutionary loss of glucose 6-phosphate dehydrogenase (*Zwf*) in YP eliminates the primary path for production of pentose precursors of a number of critical pathways [34]. Consequently, any perturbation of the alternate (and non-redundant) pathways for production of these compounds can halt cellular growth.

Finally, all SLMs from the double-knockout simulation are associated with either a) alternate pathways that produce the same critical metabolite (52%), b) a pair of critical isozymes (28%) or c) extracellular transport and reaction for *in vivo* synthesis of critical metabolites (20%). A large majority of these SLMs are associated with five critical metabolic pathways: Oxidative phosphorylation (23%), Arginine biosynthesis/urea cycle (18%), Purine biosynthesis (17%), Pentose phosphate pathway (16%) and Lysine metabolism (9%).

## Discussion

To ensure a high level of accord between publicly available experimental observations and model predictions, we had to make certain that all non-functional metabolic reactions were silenced. Given the fact that YP is a meiotrophic organism, and thus, its genome contains multiple cryptic genes, we needed to develop a method to identify these genes and correctly encode their activity in the model. Our method involves comparing single-gene knockouts for a number of select environmental conditions, extracting from this collection the few genes that can act as cryptic genes. An inspection of our results (see Table 2) shows that the model accurately identifies known genes responsible for a phenotypic behavior among the short list of candidate genes in all tested cases. Use of our methodology could serve as a powerful tool in enhancing the process of genome annotation, development of theoretical models, and design of experiments.

Gene reduction analysis of YP has shown that its genes are being inactivated or deleted under selective pressure, and that this process is likely related to the interaction of the bacterium with the nutrient rich medium in the host [51]. We used the genome-scale model to study the metabolic characteristics and cellular robustness of YP. We note that its metabolic needs differ significantly from those of its chemoheterotrophic progenitor, *Yersinia pseudotuberculosis*, as well as other members of *yersiniae* family. Examinations of the genome of YPS have shown that this organism has a full complement of the biosynthetic metabolic pathways [52]. However, an examination of our SGKO and SLM results corroborate the assertion that, compared to YPS, YP is more dependent on the host medium to provide its nutritional requirements. Some of the ubiquitous critical knockouts of YP are associated with import of obligatory nutrients, such as phenylalanine and methionine whose import from the host medium are not need by YPS.

As is characteristic of parasitic relationships, it has been noted that the metabolism of YP is very inefficient [43]. Our model simulations (see Figure 2 and Table 3) also point to a wasteful metabolism where more than half of the carbons that are imported into the cell are excreted as waste material. The predicted results show that respiratory metabolism is more carbon efficient than anoxic metabolism. Additionally, oxidative metabolism produces lower amounts of acidic byproducts, which are known to inhibit cellular growth. Because of YP's aversion to acidity, the model's results agree with the experimental practice of using gluconate as the carbon source for the batch cultivation of YP.

## Conclusions

Bubonic plague is one of the world's most dangerous diseases and still afflicts thousands of people worldwide. The discoveries of antibiotic resistant strains of *Yersinia pestis* and an ever-present threat of plague being used as a weapon of bioterrorism are added impetuses for studying the physiology of this organism and developing new therapeutics. To aid in this effort, we have developed a highly curated genome-scale constraint-based model of metabolism in YP. Concomitantly, we have developed a novel theoretical method for identification of cryptic and pseudo-genes which greatly aids the process of model development and gene annotation.

These theoretical tools will allow for faster, more accurate discovery of drug targets by identifying the cell's metabolic weaknesses and strengths. Such advances are always welcome, but in case of extremely deadly pathogens like YP, these tools are indispensable since experimental analyses are associated with the need for high-security laboratory conditions, and in some cases prohibited.

As an initial use of the model, we have developed new techniques for combining mRNA expression data with FBA models (A. Navid, E. Almaas, unpublished work) and have applied the new technique to study the YP's metabolic adaptation following its transition from flea gut environment to that of the mammalian host. Additionally we have analyzed YP's metabolic dynamics after interaction with the antibiotics Chloramphenical and Streptomycin. In the future we plan to use this model to study the metabolic capabilities of different biovars of YP as well as their progenitor *Y. pseudotuberculosis*.

# Methods

## Reconstruction of the metabolic network

Our *Yersinia pestis* model iAN818m is based on the annotated genome of strain 91001 [18] that has been deposited on the KEGG website. Our model was extensively hand-curated to ensure compliance with experimental observations, accounting for the activity of 818 of the 1146 metabolism related genes (71%) in the genome. The translated proteins from these genes catalyze 969 metabolic reactions.

From an extensive literature search and analysis of observed metabolic capabilities, we added 37 reactions that are associated with orphan enzymes (see Table 1) in addition to 14 non-enzymatic reactions. When there was a conflict between experimental observation and genomic annotation, the experimental observation took precedence. A list of characteristics of the model is given in Table 1, and a complete list of the reactions has been included in the Supplementary Material.

The biomass equation was developed using a variety of data sources: The amino acid[53] and phospholipid [54-56] composition of the biomass are unique to YP. We assumed that the nucleotide composition and the maintenance ATP usage are similar to that of YP's close relative (89% DNA sequence similarity) *Escherichia coli* [57] and is set to 7.6. A variation of this value has negligible impact on the calculated carbon efficiency (Supplementary Material). The composition for the remaining cofactors and small molecules are taken from the reported composition of prokaryotes [58].

The primary growth-determining constraint placed on the import of nutrients is an upper limit for the import of the obligate amino acid methionine. The metabolism of this nutrient is simple, i.e., it is imported and directly incorporated into the biomass. By setting an appropriate upper limit on the import of this nutrient we ensure that the growth of YP agrees with observed doubling times. Additionally, we blocked the

import of $NH_4^+$ based on observations that most strains of YP fail to show significant growth in presence of $NH_4Cl$ as the sole source of nitrogen [33], the ammonium transport facilitator gene (*amtB*) is strongly down-regulated at 37ºC [59], and finally, the fact that concentration of this compound is very small in the mammalian host. Two distinct culture media were used for all our simulations. The "minimal" medium contained gluconate, phosphate, glycine/threonine, phenylalanine, methionine, and thiosulfate. For a nutritionally rich environment which would be ideal for cellular growth, we used the TMH medium [42].

### Flux Balance Analysis (FBA)

FBA is based on representing all known metabolic reactions of an organisms by the stoichiometric matrix, $S\ (m \times n)$, where $m$ is the number of metabolites and n the number of different reactions. Applying the assumptions of mass balance and metabolic steady-state, we find the following set of linear equations governing the system's behavior:

$$\frac{dX_i}{dt} = \sum_j S_{ij} v_j = 0,$$

where $X_i$ is the concentration of metabolite $i$. Other limitations that are imposed on a system based on experimental studies enforce that the amount of flux through a reaction, the amount of nutrients imported, or waste products excreted from the organism have a lower and upper boundary:

$$\alpha \le v_i \le \beta,$$

$$\chi \le b_i \le \varphi,$$

where $v_i$ and $b_i$ are the internal and export/import flux of species $i$ respectively, and $\alpha$, $\beta$, $\chi$, and $\varphi$ are the lower and upper limit for these fluxes. Finally, FBA utilizes linear programming to determine a feasible steady-state flux vector that optimizes an objective function, most commonly chosen to be the production of biomass, i.e. cellular growth.

### CryptFind: Identification of potential cryptic/pseudo genes

The method for identifying cryptic and pseudo genes is based on the ability of FBA to predict a gene's knockout phenotype with high fidelity in quality metabolic reconstructions [60]. We generate the list of candidate cryptic genes associated with a meiotrophic trait by the following two-step process: (1) We initially identify all genes that are conditionally essential on a minimal medium only containing the nutrient source for which meiotrophy is observed. (2) We reduce this list by removing genes that are conditionally essential for any other nutrient source. The remaining candidates (if any) comprise the final list of cryptic gene candidates. For example, we identify possible cryptic genes responsible for YP's inability to ferment rhamnose by simulating system-wide single-gene knockouts in a MIN medium using rhamnose as the primary sugar source instead of gluconate. We reduce this initial list of cryptic gene candidates by removing the genes that are identified as essential through knockout analyses for a large array of other possible sugar sources. For a more detailed description of this methodology see the included supplementary material.

## Acknowledgements

We thank Dr. E. Branscomb, P. Chain, Dr. C-M. Ghim and Dr. E. Garcia for discussions. Work performed under the auspices of the U.S. Department of Energy by Lawrence Livermore National Laboratory under Contract DE-AC52-07NA27344. The project (06-ERD-061) was funded by the Laboratory Directed Research and Development program at LLNL.

## Figures

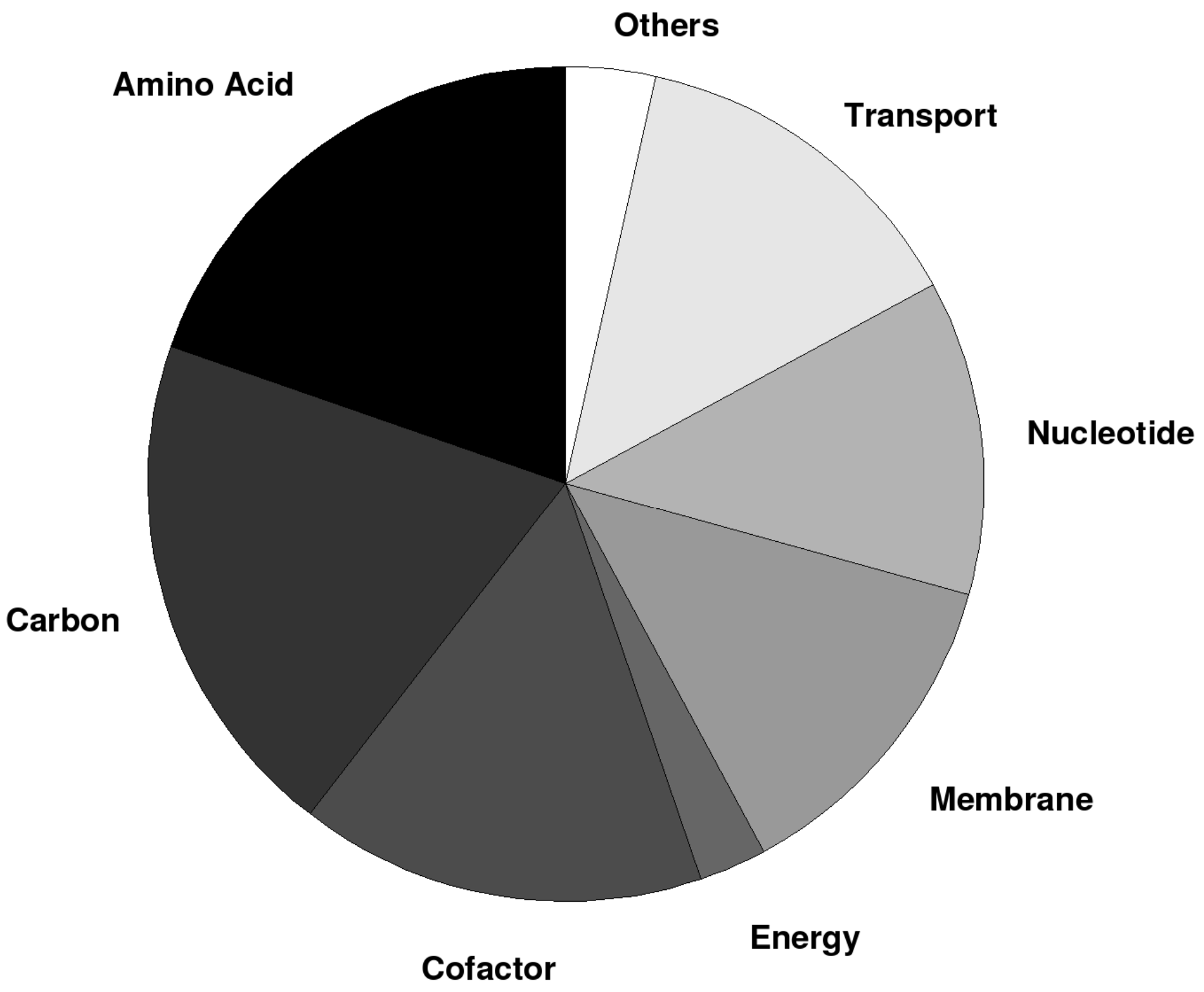

**Figure 1 - Grouping of metabolic reactions according to pathways.**

The *Yersinia pestis* strain Mediaevalis metabolic network reconstruction consists of 1020 reactions, separated into the following pathways: 200 Amino Acid, 201 Carbon, 159 Cofactor, 27 Energy, 134 Membrane, 123 Nucleotide, 142 Transport, and 34 in other pathways.

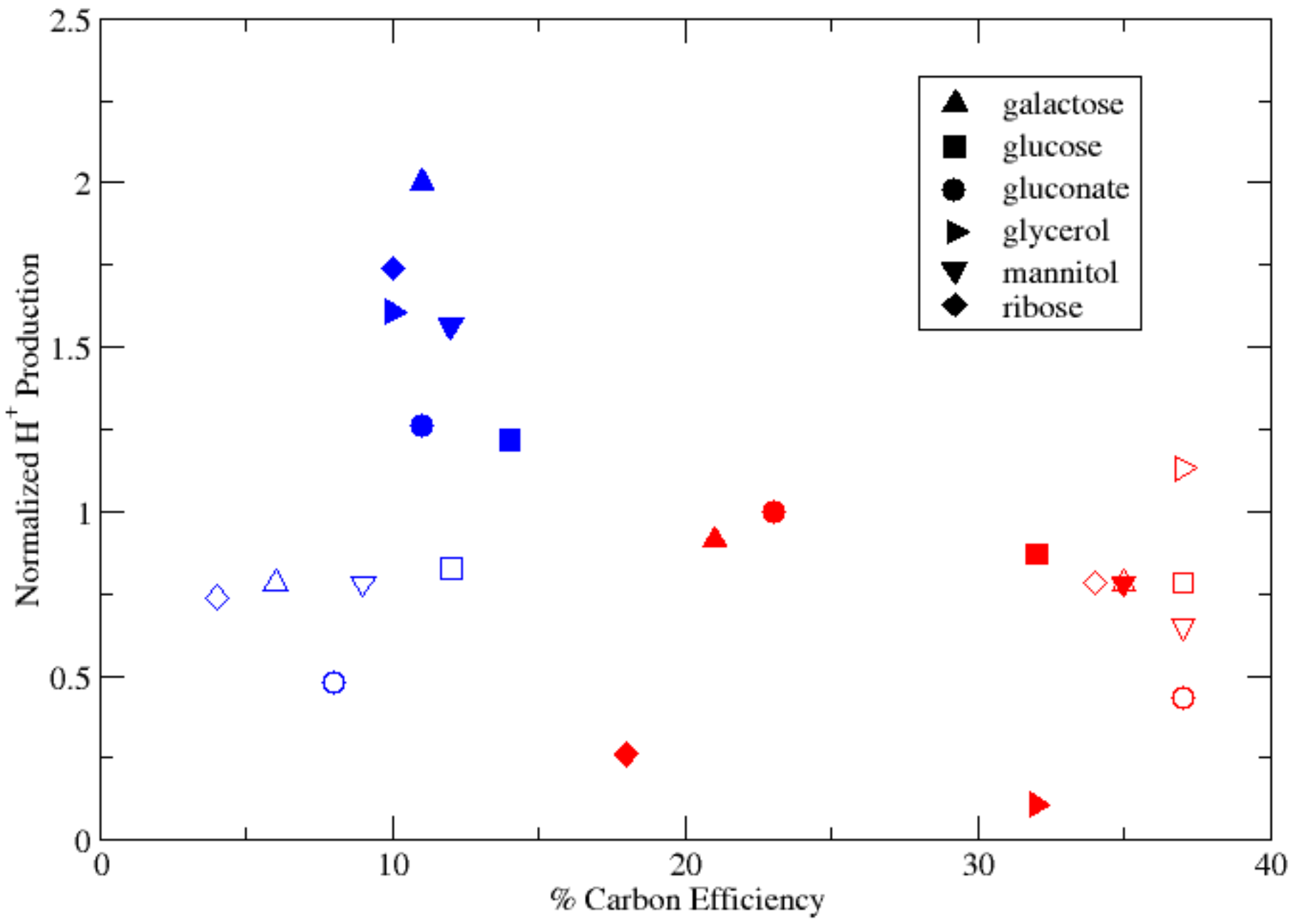

**Figure 2 - Metabolic efficiency of Yesinia pestis in varied nutritional conditions.** The calculated efficiency of carbon absorption by *Y. pestis* for a variety of carbohydrates corroborates the experimental observations[43] that the metabolism of this bacterium is wasteful. Red (blue) symbols indicate aerobic (anaerobic) conditions. Cell growth in nutrient rich or poor media is marked by filled or empty symbols, respectively. $H^+$ production is normalized to that in glucose aerobic medium. Carbon efficiency is defined as the ratio of imported carbons from the specified source to carbons incorporated in biomass.

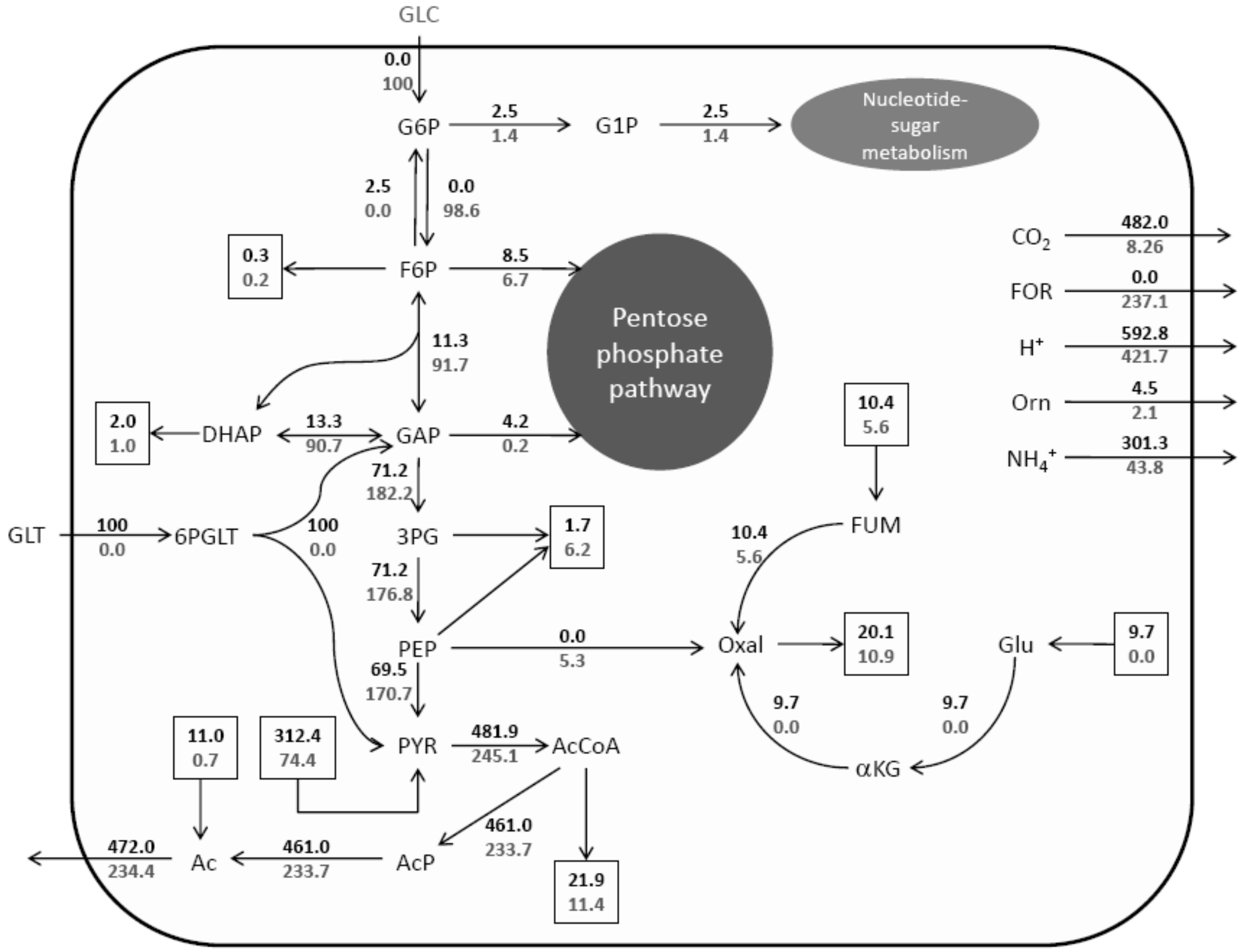

**Figure 3 - Metabolic activity of central carbon metabolism in *Y. pestis* biovar Mediaevalis**

Calculated reaction fluxes under aerobic TMH conditions with either glucose (GLC) or gluconate (GLT) as single carbon source. The calculated fluxes are given as percentage of import flux for GLT (GLC) as upper (lower) value. In agreement with Figure 2, the predicted results point to a wasteful metabolism. The results also support the reported absence or deficiency of α-ketoglutarate dehydrogenase activity in *Y. pestis*[45]. Metabolite abbreviations are G6P – glucose 6-phosphate; F6P – fructose 6-phosphate; GAP – glyceraldehyde 3-phosphate; DHAP – dihydroxyacetone phosphate; 6PGLT – 6-phopsho-gluconate; 3PG – 3-phosphoglycerate; PEP – phosphoenolpyruvate; PYR – pyruvate; AcCoA – acetyl-CoA; AcP – acetyl phosphate; Ac – acetate; Oxal – oxaloacetate; FUM – fumarate; αKG – α-ketoglutarate; Glu – glutamate; FOR – formate; Orn – ornithine.

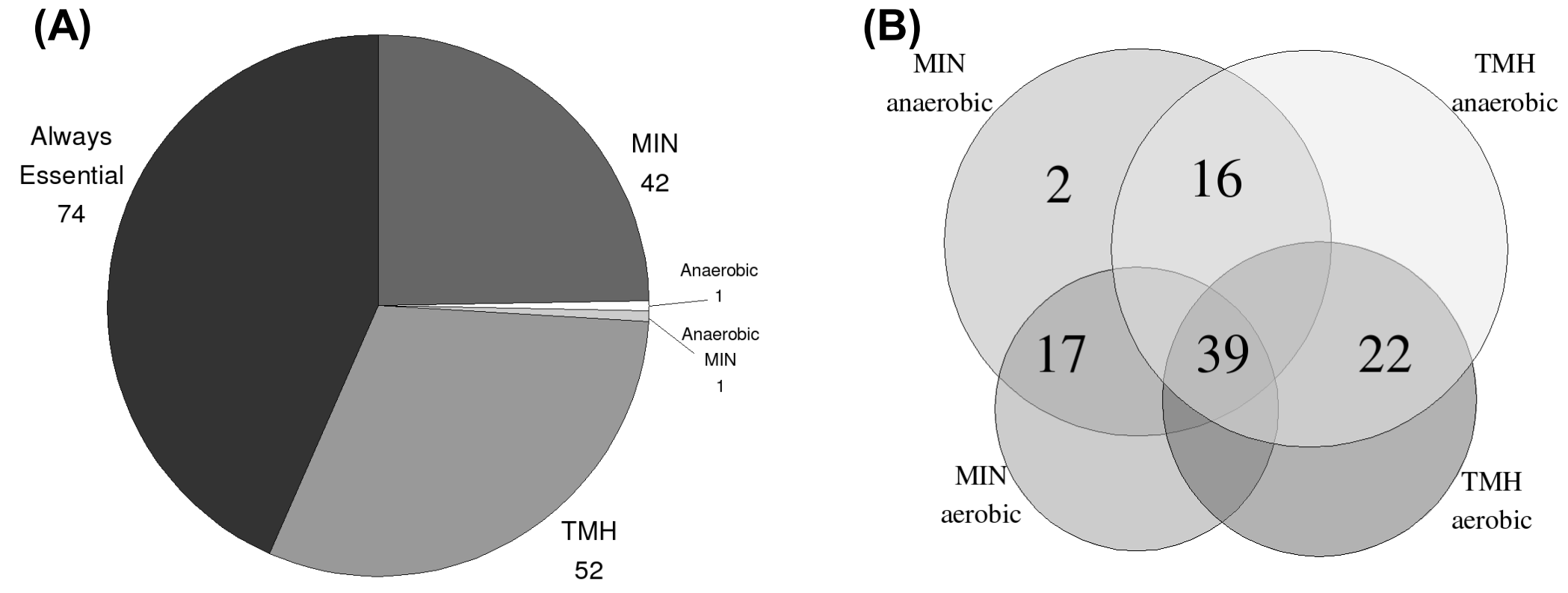

**Figure 4 -Cellular robustness to genetic mutations.**

Statistics of a) single gene knockouts (SGKO) and b) synthetic lethal mutations (SLMs) for different environments (all mixtures of aerobic, anaerobic, nutrient poor, nutrient rich). As expected, nutritionally poor media significantly limit the number of ways the metabolic network can produce necessary cellular components, and thus, cells growing in these environments are more susceptible to the deleterious effects of genetic mutations.

# Tables

**Table 1 - Model Statistics**

| | |
|---|---|
| Coding sequences | 4037 |
| DNA processing | 663 |
| Cellular processes | 825 |
| Poorly characterized | 1263 |
| Pseudogenes | 141 |
| Metabolism | 1146 |
| Genes in the model | 818 |
| Number of metabolites | 825 |
| Number of reactions | 1020 |
| -Annotated genome | 969 |
| -Non-enzymatic | 14 |
| -Literature | 20 |
| -Necessary for growth | 13 |
| -Metabolic hole filling | 4 |

**Table 2 - List of candidate cryptic/pseudo genes for *Y. pestis***

| Metabolic loss | Possible Cryptic Genes | Reactions |
|---|---|---|
| Biosynthesis of gly/thr | *metL*<br><br>*thrB*<br>*thrC* | HOM-L + $NADP^+$ ↔ ASPSA + $H^+$ + NADPH<br>ASP-L + ATP ↔ 4PASP + ADP<br>ATP + HOM-L → ADP + $H^+$ + PHOM<br>$H_2O$ + PHTHR → 4HTHR + Pi<br>$H_2O$ + PHOM → Pi+ THR-L |
| Biosynthesis of val/ile | *ilvC*<br><br>*ilvD* | 2AHBUT + $H^+$ + NADPH→ 23DHMP + $NADP^+$<br>ALAC-S + $H^+$ + NADPH →23DHMB + $NADP^+$<br>23DHMP →3MOP + $H_2O$<br>23DHMB → 3MOB + $H_2O$ |
| Biosynthesis of phe | ***pheA***[*] | AROG-L+$H^+$ ↔PHE-L+$H_2O$+$CO_2$ |
| Biosynthesis of met | *metA*<br>***metB***[*]<br><br><br><br><br><br>*metC*<br><br><br><br>*metH* | HOM-L + SUCCOA → COA + SUCHMS<br>CYS-L + SUCHMS → CYST-L + H + SUCC<br>CYS-L + ACTHMS → CYST-L + 2$H^+$ + AC<br>SUCHMS+$H_2O$ ↔2OBUT+ SUCC+ $NH_4^+$+ $H^+$<br>SUCHMS+DMS+3$H^+$ ↔ HCYS-L +SUCC+ 2$CH_4$<br>ACTHMS+DMS+3$H^+$ ↔ HCYS-L +AC+ 2$CH_4$<br>ACTHMS+SECYS →SECYST+AC+$H^+$<br>CYS-L + $H_2O$ → $H_2S$ + $NH_4^+$ + PYR<br>CYST-L + $H_2O$ → HCYS-L + $NH_4^+$ + PYR<br>CYSTI-L+$H_2O$ →PYR+$NH_4^+$+TCYS<br>SECYST+$H_2O$ →SEHCYS+$NH_4^+$+PYR<br>5MTHF + HCYS-L → MET-L + THF |
| Fermentation of arabinose | **_ara_**_AB_**_C_**[*]_FGH_ | ARAB-$L_{OUT}$+$ATP_{CYTOSOL}$+$H_2O$→<br>$ADP_{CYTOSOL}$+ARAB-$L_{CYTOSOL}$+$H^+$+Pi |
| Fermentation of rhamnose | ***betB***[†]<br><br><br>*rhaA*<br>*rhaT*<br>*tpiA* | $H_2O$ + LADL-L + $NAD^+$ → 2$H^+$ + LAC-L + NADH<br>BETALD + $H_2O$ + $NAD^+$ → GLYB + 2$H^+$ + NADH<br>BETALD + $H_2O$ + $NADP^+$ → GLYB + 2$H^+$ + NADPH<br>RMN ↔ RML<br>$H^+_{OUT}$ + $RMN_{OUT}$ → $H^+_{CYTOSOL}$ + $RMN_{CYTOSOL}$<br>DHAP ↔ G3P |
| Fermentation of melibiose | *glk*<br>***melB***[*] | ATP + GLC-D → ADP + G6P + $H^+$<br>$H^+_{OUT}$ + $MELIB_{OUT}$ → $H^+_{CYTOSOL}$ + $MELIB_{CYTOSOL}$ |
| Urease | *YPO2672*<br>**_ure_**_ABC_**_D_**[*]_EFG_ | $UREA_{OUT}$ ↔ $UREA_{CYTOSOL}$<br>UREA+$H_2O$+ 2$H^+$ →$CO_2$+ 2$NH_4^+$ |

[*]known cryptic/pseudo genes
[†]likely cryptic gene

**Table 3 -Comparative analysis of optimal growth of *Y. pestis* utilizing different carbon sources.**

| Carbohydrate | Normalized Growth | Normalized $CO_2$ export | P/O ratio | Normalized $H^+$ production |
|---|---|---|---|---|
| Aerobic MIN | | | | |
| galactose | 1.0 | 0.6 | 1.3 | 1.8 |
| glucose | 1.0 | 0.4 | 1.2 | 1.8 |
| gluconate | 1.0 | 1.0 | 1.3 | 1.0 |
| glycerol | 1.0 | 0.4 | 0.8 | 2.6 |
| mannitol | 1.0 | 0.4 | 1.1 | 1.5 |
| ribose | 1.0 | 0.6 | 1.3 | 1.8 |
| Anaerobic MIN | | | | |
| galactose | 0.1 | -0.5 | N/A | 1.8 |
| glucose | 0.3 | -0.5 | N/A | 1.9 |
| gluconate | 0.2 | -0.2 | N/A | 1.1 |
| mannitol | 0.2 | -0.6 | N/A | 1.8 |
| ribose | 0.1 | -0.5 | N/A | 1.7 |
| Aerobic TMH | | | | |
| galactose | 1.0 | 2.1 | 1.3 | 2.1 |
| glucose | 1.0 | 0.0 | 1.2 | 2.0 |
| gluconate | 1.0 | 0.1 | 1.2 | 2.3 |
| glycerol | 1.0 | 0.4 | 1.1 | 0.3 |
| mannitol | 1.0 | 0.0 | 1.3 | 1.8 |
| ribose | 1.0 | 2.0 | 1.3 | 0.6 |
| Anerobic TMH | | | | |
| galactose | 1.0 | -0.8 | N/A | 4.6 |
| glucose | 1.0 | 0.0 | N/A | 2.8 |
| gluconate | 1.0 | 0.0 | N/A | 2.9 |
| mannitol | 1.0 | 0.0 | N/A | 3.6 |
| ribose | 1.0 | 0.0 | N/A | 4.0 |